\documentclass[pra,twocolumn,floatfix,amsmath,amssymb]{revtex4-1}
\usepackage{physics}
\usepackage{graphicx}
\usepackage{mathdots}
\usepackage{bm}
\usepackage{transparent}
\usepackage{times}
\usepackage[scaled=.9]{helvet}
\usepackage[usenames,dvipsnames]{xcolor}
\usepackage{hyperref}
\hypersetup{
	colorlinks,
	allcolors=NavyBlue,
	linktoc=all
}

\usepackage[capitalize]{cleveref}
\crefformat{section}{Sec.~#2#1#3}

\renewcommand{\Re}{\mathrm{Re}}
\renewcommand{\Im}{\mathrm{Im}}
\newcommand{\cc}{\mathcal C}
\newcommand{\G}{\mathcal G}	
 
\newcommand*\dagg{^{\dagger}}
\newcommand*\eps{\varepsilon}
\newcommand{\D}{\mathcal{D}}
\newcommand*\dens{{\rho}}  
\newcommand*\mat[1]{\begin{pmatrix}#1\end{pmatrix}} 
\newcommand*\matr[1]{\bm{#1}}
\graphicspath{{figs/}}

\begin{document}

\title{Optomechanical dual-beam backaction-evading measurement\\ beyond the rotating-wave approximation}
\author{Daniel Malz}
\author{Andreas Nunnenkamp}
\affiliation{Cavendish Laboratory, University of Cambridge, Cambridge CB3 0HE, United Kingdom}
\date{\today}

\begin{abstract}
	We present the exact analytical solution of the explicitly time-periodic quantum Langevin equation describing the dual-beam backaction-evasion measurement of a single mechanical oscillator quadrature due to Braginsky, Vorontsov and Thorne beyond the commonly used rotating-wave approximation.
	We show that counterrotating terms lead to extra sidebands in the optical and mechanical spectra and to a modification of the main peak.
	Physically, the backaction of the measurement is due to periodic coupling of the mechanical resonator to a light field quadrature that only contains cavity-filtered shot noise.
	Since this fact is independent of other degrees of freedom the resonator might be coupled to, our solution can be generalized,
	including to dissipatively or parametrically squeezed oscillators, as well as recent two-mode backaction evasion measurements.
\end{abstract}

\maketitle

\section{Introduction}
A continuous measurement of the position of a harmonic oscillator is subject to the ``standard quantum limit'' (SQL),
a limit directly imposed by Heisenberg's uncertainty relation~\cite{Caves1980,Clerk2010}.
An observable that can be monitored without precision limit is called ``quantum non-demolition'' (QND) variable,
such that its continuous measurement can avoid the measurement backaction (BA)~\cite{Braginsky1980}
and thus open the way to the detection of weak forces, such as those due to gravitational waves~\cite{Abbott2016}.

There has been continued interest to implement backaction-evading (BAE) measurements~\cite{Bocko1996}.
Following a detailed theoretical proposal, the first demonstration with a sensitivity beyond the SQL was in optomechanics~\cite{Clerk2008,Hertzberg2010}
and they have since proven very useful~\cite{Suh2014,Lecocq2015,Lei2016,Polzik2014}.
Despite the importance of such measurements, Ref.~\cite{Clerk2008} discusses only lowest-order corrections to the rotating-wave approximation (RWA).

Here, using the recently developed Floquet approach~\cite{Malz2016}, we derive the exact solution to the equations describing a BAE measurement.
Due to the presence of CR terms, this constitutes a solution to genuinely explicitly time-dependent quantum Langevin equations, as there is no frame in which they become stationary.
The solution is possible because the mechanical oscillator couples solely to a light quadrature that is independent of the mechanical oscillator, and only contains filtered shot noise.
The coupling is periodic, which  leads to two-time correlators that are not time-translation invariant.
In such a situation, the power spectrum is a time average of the Fourier transform of the autocorrelator~\cite{Malz2016}.

In the following, after introducing our model in \cref{sec:model}, we briefly remind the reader of results in RWA in \cref{sec:RWA}.
Then, we present all aspects of the solution, from the Floquet framework to the resulting spectra and backaction in \cref{sec:framework}.
Finally, in \cref{sec:general recipe}, we show how to generalize our solution to the important cases of dissipative and parametric squeezing, as well as two-mode BAE measurements
(with explicit calculations in \cref{app:two-mode,app:dual-beam squeezing,app:weak coupling,app:parametric squeezing}).
We conclude in \cref{sec:conclusion}.

\section{Model}\label{sec:model}
\begin{figure}[t]
  \centering
  \includegraphics[width=\linewidth]{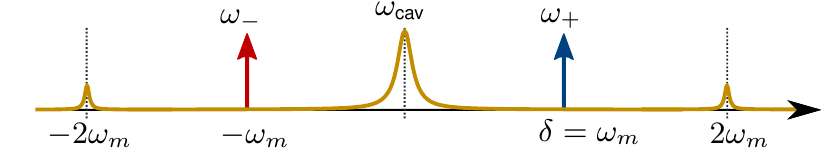}
  \caption{The dual-beam backaction-evading (BAE) measurement scheme proposed in Ref.~\cite{Braginsky1980,Clerk2008}.
  Two lasers of equal strength drive the cavity at frequencies $\omega_\pm=\omega_{\text{cav}}\pm\delta$, where $\delta=\omega_m$ such that their sidebands overlap at the cavity frequency $\omega_{\text{cav}}$.
  Counterrotating (CR) terms cause peaks at $\omega_{\text{cav}}\pm2\omega_m$ to appear.
  The output light will be an approximate BAE measurement of a rotating quadrature of the mechanical oscillator,
  the CR terms being responsible for the finite backaction.}
  \label{fig:driving scheme}
\end{figure}
In cavity optomechanics, the photons in a cavity couple to the motion of a mirror via radiation pressure.
A BAE measurement of one of the quadratures of the oscillator can be implemented by applying two drives of equal strength at frequencies $\omega_{\text{cav}}\pm\omega_m$ (see \cref{fig:driving scheme}).
Here, $\omega_{\text{cav}} (\omega_m)$ is the frequency of a cavity (mechanical) mode.
Originally due to Braginsky et al.~\cite{Braginsky1980}, the first detailed analysis in the context of cavity optomechanics was given in Ref.~\cite{Clerk2008}.

We consider the Hamiltonian of a standard cavity optomechanical system
\begin{equation}
  	H=H_{\text{OM}}+H_{\text{rest}}+H_{\text{drives}}+H_{\text{baths}},
\end{equation}
where ($\hbar=1$)
\begin{subequations}
	\begin{align}
		H_{\text{OM}}&=\omega_{\text{cav}} a\dagg a+\omega_m b\dagg b-g_0 a\dagg a( b\dagg+ b),\\
		H_{\text{drives}}&=\alpha(e^{-i\omega_+t}+e^{-i\omega_-t}) a\dagg+\text{h.c.}
	\end{align}
\end{subequations}
$a, b$ are the bosonic annihilation operators of the cavity mode and the mechanical oscillator, respectively.
The cavity mode frequency is $\omega_{\text{cav}}$,
the mechanical frequency $\omega_m$, the coupling strength via radiation pressure $g_0$,
and the driving strength of the drives with frequencies $\omega_\pm=\omega_{\text{cav}}\pm\delta$ is $\alpha$, 
where we have left $\delta$ unspecified for now (\cref{fig:driving scheme}).
The BAE measurement is realized for $\delta=\omega_m$.
A detailed derivation of the optomechanical Hamiltonian can be found for instance in Ref.~\cite{Aspelmeyer2014}.
$H_{\text{baths}}$ couples the oscillator and cavity modes to baths at finite and zero temperature, respectively.
$H_{\text{rest}}$ will remain unspecified for now, it can contain terms that couple the harmonic oscillator to other degrees of freedom.
The only requirement is that $[H_{\text{rest}}, a+a\dagg]=0$, i.e., that the other degrees of freedom do not couple to this measurement cavity quadrature.

To proceed, we split the light field into a coherent part and fluctuations,
go into a rotating frame, $a=\bar ae^{-i\omega_{\text{cav}}t}(e^{i\delta t}+e^{-i\delta t}+ d)$,
and linearize the Hamiltonian. Without loss of generality, $\delta>0$.
Under the usual assumptions of Markovian baths and a one-sided cavity, the resulting Hamiltonian
\begin{equation}
  \label{eq:lin Hamiltonian}
  	H=\omega_m b\dagg b-2G\cos(\delta t)( b+ b\dagg)( d+ d\dagg)
\end{equation}
gives rise to Langevin equations~\cite{gardiner2004quantum,Clerk2010} that are periodic in time
\begin{subequations}
  \begin{align}
  	\dot{ d}&=-\frac{\kappa}{2} d+\sqrt{\kappa} d_{\text{in}}
  	+2iG\cos(\delta t)( b+ b\dagg),
  	\label{eq:langevin1}\\
  	\dot{ b}&=\left(-i\omega_m-\frac{\gamma}{2}\right) b+\sqrt{\gamma} b_{\text{in}}\nonumber\\
  	&\qquad	+2iG\cos(\delta t)( d+ d\dagg)+i[H_{\text{rest}}, b].
  	\label{eq:langevin2}
  \end{align}
\end{subequations}
Here, we have defined the enhanced optomechanical coupling constant $G=g_0\bar a$, and the mechanical (optical) dissipation rate $\gamma$ ($\kappa$).
$b_{\text{in}}, d_{\text{in}}$ are input noise operators with
$\langle  d_{\text{in}}(t) d_{\text{in}}\dagg(t')\rangle=\delta(t-t')$,
$\langle  d_{\text{in}}\dagg(t) d_{\text{in}}(t')\rangle=0$,
$\langle  b_{\text{in}}(t) b_{\text{in}}\dagg(t')\rangle=(n_{\text{th}}+1)\delta(t-t')$, and
$\langle  b_{\text{in}}\dagg(t) b_{\text{in}}(t')\rangle=n_{\text{th}}\delta(t-t')$.

In \cref{sec:solution}, we solve \cref{eq:langevin1,eq:langevin2} without further approximations,
in particular without the rotating wave approximation (RWA).

\section{Backaction evasion in RWA}\label{sec:RWA}
As a reminder, we first consider a BAE measurement within RWA.
We define a quadrature rotating a frequency $\delta$ as 
\begin{equation}
  X_\delta=be^{i\delta t}+b\dagg e^{-i\delta t}.
  \label{eq:rotating quadrature}
\end{equation}
The cavity equation of motion is (cf.~\cref{eq:langevin1})
\begin{equation}
	\dot d=-\frac{\kappa}{2}d+\sqrt{\kappa}d_{\text{in}}+iG\left(be^{i\delta t}+b\dagg e^{-i\delta t}\right).
	\label{eq:d eom in RWA}
\end{equation}
The cavity couples only to one quadrature of the mechanical oscillator $X_\delta$ (set by the phase relation of the external drives).
The equation of motion for that quadrature can be obtained from \cref{eq:langevin2} (here also in RWA)
\begin{multline}
	\dot{X_\delta}=-\frac{\gamma}{2}X_\delta+\sqrt{\gamma}X_{\delta,\text{in}}+i[H_{\text{rest}},X_\delta]\\
	+iG\left[ d\left(e^{i(\omega_m-\delta)t}-e^{i(\delta-\omega_m)t} \right)-\text{h.c.} \right].
\end{multline}
If $\delta=\omega_m$, the term in square brackets in the second line vanishes,
and $X_{\omega_m}$ is entirely unaffected by the cavity.
We can readily solve the equation of motion for $d$, \cref{eq:d eom in RWA}
\begin{equation}
	\label{eq:readout optics annihilation operator}
	d(\omega)=\chi_c(\omega)\left[ \sqrt{\kappa}d_{\text{in}}(\omega)+iGX_\delta(\omega) \right],
\end{equation}
with cavity response function $\chi_c(\omega)=(\kappa/2-i\omega)^{-1}$.
Thus the optical spectrum is directly related to the quadrature spectrum
\begin{equation}
	\label{eq:ideal readout spectrum}
	S_{d\dagg d}(\omega)=|\chi_c(\omega)|^2G^2S_{X_\delta X_\delta}(\omega),
\end{equation}
where for a stationary process the Wiener-Khinchin theorem ensures that the Fourier transform
\begin{equation}
	S_{AB}(\omega)=\int_{-\infty}^\infty\dd{t}e^{i\omega t}\ev{A(t)B(0)}
	\label{eq:stationary spectrum definition}
\end{equation}
is the quantum noise spectral density.
The scheme corresponds to a true BAE measurement, if the mechanical quadrature rotating with frequency $\delta$ is a QND variable.
Here, this is the case for $\delta=\omega_m$ (or, if applicable, $\delta=\omega_{m,\text{eff}}$, the effective mechanical frequency).
Independent of RWA, the input-output relation $d_{\text{out}}=d_{\text{in}}-\sqrt{\kappa}d$ can be used to the obtain the output spectrum
$S_{d_{\text{out}}\dagg d_{\text{out}}}(\omega)=\kappa S_{d\dagg d}(\omega).$
For details on how to measure and interpret the output spectrum, see Ref.~\cite{Weinstein2014}.

\section{Floquet framework}\label{sec:framework}
The analysis in \cref{sec:RWA} above breaks down if we include counterrotating (CR) terms.
Using the framework developed in Ref.~\cite{Malz2016}, it is possible to find the stationary state of the explicitly time-periodic quantum Langevin equations exactly, which
gives us the opportunity to precisely determine the backaction of the BAE measurement.

First, we split system operators into Fourier components
\begin{subequations}
  	  \label{eq:fourier split}
	\begin{align}
  		d(t)&=\sum_{n=-\infty}^{\infty}e^{in\delta t}d^{(n)}(t),\\
  		d\dagg(t)&=\sum_{n=-\infty}^{\infty}e^{in\delta t}d^{(n)\dag}(t),
	\end{align}
\end{subequations}
and define
\begin{subequations}
	\begin{align}
  		d^{(n)}    (\omega)&=\int_{-\infty}^\infty\dd{t}e^{i\omega t}d^{(n)}(t),\\
  		d^{(n)\dag}(\omega)&=\int_{-\infty}^\infty\dd{t}e^{i\omega t}d^{(n)\dag}(t).
	\end{align}
\end{subequations}
Note that this convention results in $[d^{(n)}(\omega)]\dagg=d^{(-n)\dag}(-\omega)$.
The reason for doing this is that \cref{eq:langevin1,eq:langevin2} can now be written as an infinite set of coupled,
\emph{stationary} Langevin equations
\begin{subequations}
\begin{multline}\label{eq:fourier1}
	\dot d^{(n)}=\left(-in\delta-\frac{\kappa}{2}\right)d^{(n)}+\delta_{n,0}\sqrt{\kappa}d_{\text{in}}\\
	+iG\left( b^{(n-1)}+b^{(n+1)\dag}+\lambda b^{(n+1)}+\lambda b^{(n-1)\dag} \right),
\end{multline}
\begin{multline}\label{eq:fourier2}
	\dot b^{(n)}=\left( -i\omega_m-in\delta-\frac{\gamma}{2} \right)b^{(n)}+\delta_{n,0}\sqrt{\gamma}b_{\text{in}}\\
	+iG\left(I^{(n+1)}+ \lambda I^{(n-1)} \right)+i[H_{\text{rest}}, b]^{(n)}.
\end{multline}
\end{subequations}
where $x^{(n)}\equiv b^{(n)}+b^{(n)\dag}$, $I^{(n)}\equiv d^{(n)}+ d^{(n)\dag}$, $\delta_{n,0}$ is the Kronecker delta, and $[H_{\text{rest}},b]^{(n)}$ is the $n$th Fourier component of the commutator.
Where feasible, here and in the following, we will mark counterrotating (i.e., off-resonance) terms by a $\lambda$ (note that this only makes sense when $\delta$ is close to $\omega_m$),
such that RWA corresponds to $\lambda=0$ and the full solution to $\lambda=1$.
To further guide the intuition, we remark that the second line of \cref{eq:fourier1} equals
$iG(x^{(n-1)}+x^{(n+1)})$. One can think of the two laser drives to result in two separate couplings to the position $x$ of the resonator.

We define the ``spectrum''
\begin{equation}
  S_{A\dagg A}(\omega,t)\equiv\int_{-\infty}^\infty\dd{\tau}e^{i\omega\tau}C_{AA}(\tau,t),
  \label{eq:spectrum}
\end{equation}
with $C_{AA}(\tau,t)\equiv\ev{A\dagg(t+\tau)A(t)}$.
The time dependence of $S_{A\dagg A}(\omega,t)$ can be expressed as a Fourier series~\cite{Malz2016}
\begin{equation}
  S_{A\dagg A}(\omega,t)=\sum_{n=-\infty}^\infty e^{i\delta nt}S_{A\dagg A}^{(n)}(\omega)
  \label{eq:spectrum fourier components}
\end{equation}
with components
\begin{equation}
  S_{A\dagg A}^{(m)}(\omega)=\sum_n\int\frac{\dd{\omega'}}{2\pi}
  \ev{A^{(n)\dag}(\omega+n\delta)A^{(m-n)}(\omega')}.
  \label{eq:fourier components}
\end{equation}
It can be shown that the zeroth Fourier component, which at the same time is the time average of $S(\omega,t)$,
is the measured power spectrum~\cite{Malz2016}.

\subsection{Solution beyond RWA}\label{sec:solution}
To solve \cref{eq:langevin1,eq:langevin2}, we define the optical quadratures $I=d+d\dagg$ and $Q=-i(d-d\dagg)$.
From \cref{eq:langevin1} we obtain
\begin{subequations}
	\begin{align}
		\dot I&=-\frac{\kappa}{2}I+\sqrt{\kappa}I_{\text{in}},\label{eq:I eom}\\
		\dot Q&=-\frac{\kappa}{2}Q+\sqrt{\kappa}Q_{\text{in}}+4G\cos(\delta t)x.
	\end{align}
	\label{eq:optical quadrature eoms}
\end{subequations}
\begin{figure}[t]
	\centering
	\includegraphics[width=\linewidth]{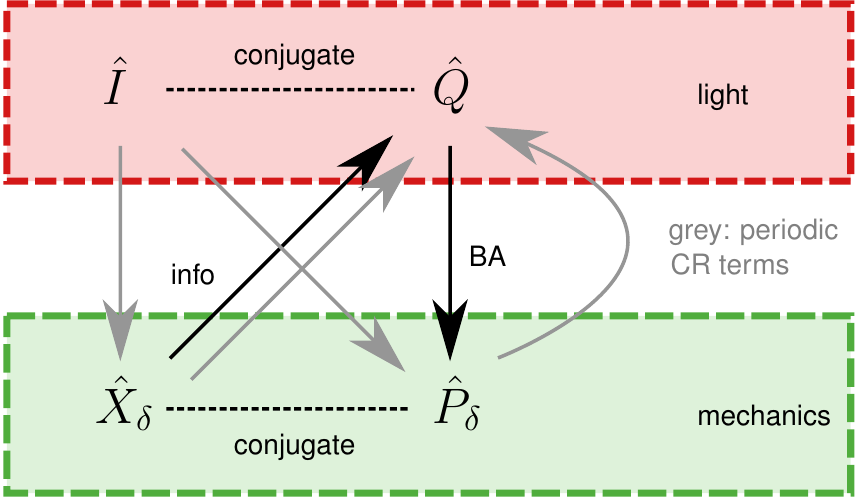}
	\caption{Coupling of quadratures. 
		$\hat Q$ records the desired information (info) about $\hat X_\delta$. 
		The measurement backaction (BA) is acting on $\hat P_\delta$.
		Marked in gray are coupling due to counterrotating (CR) terms that are neglected in RWA.}
	\label{fig:quad_diag}
\end{figure}
For an illustration how the mechanical and optical quadratures couple together, see \cref{fig:quad_diag}.
For example, the arrow from $X_\delta$ to $Q$ indicates that $X_\delta$ appears in the equation of motion of $Q$ and that the latter therefore is influenced by the former.
Since $I$ commutes with the system Hamiltonian~\eqref{eq:lin Hamiltonian}, $[H,I]=0$, there are no arrows pointing toward it in \cref{fig:quad_diag} and we can solve its equation of motion \eqref{eq:I eom} directly
\begin{equation}
	I(\omega)=\sqrt{\kappa}\chi_c(\omega)I_{\text{in}}(\omega),\quad\text{and}\quad I^{(n)}(\omega)=\delta_{n,0}I(\omega).
	\label{eq:I}
\end{equation}
Thus $\ev{I(\omega)I(\omega')}=2\pi\delta(\omega+\omega')|\chi_c(\omega)|^2$ and $\ev{I(\omega)}=0$.
\Cref{eq:I eom,eq:I} imply that $I$ is shot noise, filtered by the cavity, and is independent of the mechanics.
Furthermore, this result does not depend on $H_{\text{rest}}$, as long as $[H_{\text{rest}},I]=0$.
The coupling to $I$ is periodic (cf.\ \cref{eq:langevin2}), a consequence of amplitude beating of the coherent state in the cavity. 
Therefore, the measurement has the same effect on the mechanical resonator as a time-periodic coupling to filtered shot noise.

With \cref{eq:I} we can solve \cref{eq:fourier2} in the case $H_{\text{rest}}=0$ by going into frequency space
\begin{subequations}
	\begin{align}
		b^{(n)}(\omega)&=\chi_m(\omega-n\delta)\left[ \delta_{n,0}\sqrt{\gamma}b_{\text{in}}+2\sqrt{\kappa}Gf_{\text{in}}^{(n)}\right],\\
		b^{(n)\dag}(\omega)&=\chi_m^*(-\omega+n\delta)\left[ \delta_{n,0}\sqrt{\gamma}b_{\text{in}}\dagg-2\sqrt{\kappa}Gf_{\text{in}}^{(n)}\right],
	\end{align}
	\label{eq:mech fourier cpts}
\end{subequations}
where the new bath noise operators
\begin{equation}
	f_{\text{in}}^{(n)}(\omega)=i\chi_c(\omega)(\delta_{n,1}+\lambda\delta_{n,-1})I_{\text{in}}(\omega)/2.
	\label{eq:time-dependent bath}
\end{equation}
They obey $f_{\text{in}}^{(n)\dag}=-f_{\text{in}}^{(n)}$, $f_{\text{in}}^{(-n)}=f_{\text{in}}^{(n)}$,
and $\ev{f_{\text{in}}}=0$. The time-dependence is best seen in the time domain, where
\begin{equation}
	f_{\text{in}}(t)=i\cos(\delta t)I(t)/\sqrt{\kappa}.
	\label{eq:fin in time}
\end{equation}
This expression explicitly contains the time-dependent coupling and the filtered shot noise $I(t)$.
The correlator is
\begin{equation}
	\ev{f_{\text{in}}(t)f_{\text{in}}(t')}=-e^{-\kappa|t-t'|/2}\cos(\delta t)\cos(\delta t').
	\label{eq:fin time correlator}
\end{equation}
For stationary noise, the RHS of \cref{eq:fin time correlator} would have to depend solely on the difference $t-t'$.

We can rewrite the equation of motion for $b$ in terms of the new input $f_{\text{in}}$
\begin{equation}
	\dot b=\left( -i\omega_m-\frac{\gamma}{2} \right)b+\sqrt{\gamma}b_{\text{in}}
	+2\sqrt{\kappa}G f_{\text{in}}+i[H_{\text{rest}},b].
	\label{eq:b eom in time}
\end{equation}
From this equation it is clear that we are always able to pass from the ``unmeasured system'' ($G=0$) to the ``measured'' one ($G\neq0$) by substituting
\begin{equation}
	\sqrt{\gamma}b_{\text{in}}\to \sqrt{\gamma}b_{\text{in}}+2\sqrt{\kappa}Gf_{\text{in}}.
	\label{eq:replacement rule}
\end{equation}
We use this to generalize the solution in \cref{sec:general recipe}.

\subsection{Importance of Floquet framework}\label{sec:importance}
At this point it is useful to reflect on the advantage of the Floquet approach.
It might appear as if it were unnecessary, since with \cref{eq:I eom,eq:I}, we can already write down 
\begin{equation}
	\dot b=\left( -i\omega_m-\frac{\gamma}{2} \right)b+\sqrt{\gamma}b_{\text{in}}+2iG\cos(\delta t)I(t)+i[H_{\text{rest}},b],
\end{equation}
which for $H_{\text{rest}}=0$ has the solution
\begin{equation}
	b(\omega)=\chi_m(\omega)\left\{ \sqrt{\gamma}b_{\text{in}}(\omega)+iG\left[ I(\omega-\delta)+I(\omega+\delta) \right] \right\}.
	\label{eq:naive solution}
\end{equation}
Since $x(\omega)=b(\omega)+b\dagg(\omega)$, this might prompt us to derive the ``power spectrum''
\begin{equation}
	S_{xx}(\omega)=\int\frac{\dd\omega'}{2\pi}\ev{x(\omega)x(\omega')}.
	\label{eq:wrong power spectrum}
\end{equation}
This, however, is \emph{not} the correct power spectrum, since $x(t)$ does not describe a stationary stochastic process.
It is possible to remove the non-stationary terms manually (above they will be $\ev{I(\omega-\delta)I(\omega'-\delta)}$ and
$\ev{I(\omega+\delta)I(\omega'+\delta)}$) and thus obtain the stationary part, which \emph{is} the power spectrum~\cite{Malz2016}.
In contrast, the systematic solution via Fourier components, which all obey stationary Langevin equations, is well-defined.
In addition, the Fourier components are more versatile, and allow writing down the spectrum in arbitrary rotating frames~\cite{Malz2016}.
They also provide more intuition, since different Fourier components tend to have different physical origins.
Finally, if no exact solution is viable, they simplify the process of approximating to desired order.

\subsection{Mechanical quadrature spectrum}\label{sec:mechanical spectrum}
We define a rotating quadrature as before (see \cref{eq:rotating quadrature}).
Its spectrum is~\cite{Malz2016}
\begin{multline}
	S_{X_\delta X_\delta}^{(0)}(\omega)=S_{bb}^{(-2)}(\omega+\delta)+S_{bb\dagg}^{(0)}(\omega+\delta)\\
	+S_{b\dagg b\dagg}^{(2)}(\omega-\delta)+S_{b\dagg b}^{(0)}(\omega-\delta),
		\label{eq:rotating spectrum}
\end{multline}
where $S_{AB}$ is defined through \cref{eq:fourier components}.
We have only written down the zeroth Fourier component (the stationary part), because that is the part of the spectrum that will be measured.
We can split it up into two parts
\begin{equation}
	S_{X_\delta X_\delta}^{(0)}(\omega)=S^{\text{RWA}}_{X_\delta X_\delta}(\omega)+S^{\text{CR}}_{X_\delta X_\delta}(\omega),
\end{equation}
with $S_{X_\delta X_\delta}^{\text{RWA}}$ being the result in RWA (dependent on $H_{\text{rest}}$),
and $S_{X_\delta X_\delta}^{\text{CR}}$ being due to CR terms. Note that $S_{X_\delta X_\delta}^{\text{RWA}}$ is unchanged from the unmeasured case, since this is the essence of a BAE measurement. 
For $H_{\text{rest}}=0$, we obtain our first major result
\begin{multline}
	\label{eq:first major result}
	S_{X_{\delta}X_{\delta}}^{\text{CR}}(\omega)=\kappa \lambda^2 G^2\left( |\chi_m(\omega+\delta)\chi_c(\omega+2\delta)|^2\right.\\
	\left.+|\chi_m(-\omega+\delta)\chi_c(-\omega+2\delta)|^2 \right).
\end{multline}

If $H_{\text{rest}}$ is nonzero, but couples weakly, such that its effect is well approximated by introducing an effective
damping constant and mechanical frequency, the analysis is unchanged, such that the correction takes the same form
\begin{multline}
	S_{X_\delta X_\delta}^{\text{CR}}(\omega)
	=\kappa \lambda^2 G^2\left\{ |\chi_{m,\text{eff}}(\omega+\delta)\chi_c(\omega+2\delta)|^2\right.\\
	+\left.|\chi_{m,\text{eff}}(-\omega+\delta)\chi_c(-\omega+2\delta)|^2 \right\},
	\label{eq:CR correction}
\end{multline}
but with an effective susceptibility $\chi_{m,\text{eff}}(\omega)=[\Gamma_{\text{eff}}-i(\omega-\omega_{m,\text{eff}})]^{-1}$.
Note that if $\delta\neq\omega_{m,\text{eff}}$, the measured quadrature rotates at a different frequency than the natural oscillator quadratures
and thus the measurement backaction (BA) will contaminate the measurement at later times, even in RWA.
Only the case $\delta=\omega_{m,\text{eff}}$ is backaction-evading (BAE).
Since this is our main interest, we will fix $\delta=\omega_{m,\text{eff}}$ in the following.
For the general case $\delta\neq\omega_{m,\text{eff}}$, see \cref{eq:full quadrature spectrum}.
The change in the spectrum \cref{eq:CR correction} includes a modification of the main peak,
and new peaks corresponding to the upper sideband of the blue drive and the lower sideband of the red drive.

\subsection{Optical spectrum}\label{sec:optical spectrum}
\begin{figure}[t]
  \centering
	\includegraphics[width=\linewidth]{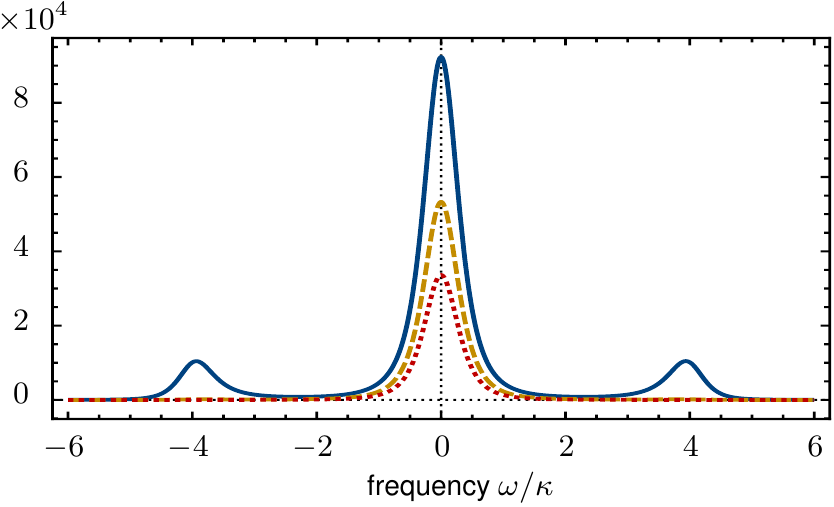}
  	\caption{
  		Exact optical output spectrum $\kappa S_{d\dagg d}^{(0)}$ \cref{eq:sdd} (blue, solid)
  		in comparison to a perfect BAE measurement of the modified mechanical spectrum \eqref{eq:CR correction} from \cref{sec:mechanical spectrum} (yellow, dashed),
  		and the RWA result \cref{eq:ideal readout spectrum} (red, dotted).
  		The full spectrum (blue, solid) is calculated via \cref{eq:sdd}, where additional terms due to the counterrotating quadrature $X_{-\delta}$ appear. 
  		The sidepeaks of the modified mechanical spectrum do not show in the yellow curve, as they are suppressed by $|\chi_c(\omega)|^2$.
  		Parameters $\Gamma_{\text{eff}}/\kappa=1,\omega_m/\kappa=2,n_{\text{eff}}=10,G/\kappa=10,\delta=\omega_m$ (thus $\cc=400$).
  		Note that the large value of $\Gamma_{\text{eff}}$ is chosen for visibility.
  	}
  	\label{fig:spectra}
\end{figure}
As we have seen, CR terms modify the mechanical quadrature spectrum, which is reflected in the optical spectrum
\begin{equation}
	\begin{aligned}
		S_{d\dagg d}^{(0)}(\omega)&=G^2|\chi_c(\omega)|^2\left[ S_{xx}^{(0)}(\omega-\delta)+S_{xx}^{(0)}(\omega+\delta)\right.\\
			&\qquad\left.+S_{xx}^{(2)}(\omega-\delta)+S_{xx}^{(-2)}(\omega+\delta)\right]\\
			&=G^2|\chi_c(\omega)|^2\left[ S_{X_\delta X_\delta}^{(0)}(\omega)+\lambda^2S_{X_{-\delta}X_{-\delta}}^{(0)}(\omega)\right.\\
				&\qquad\left.+\lambda\left( S_{X_\delta X_{-\delta}}^{(0)}(\omega)+S_{X_{-\delta}X_\delta}^{(0)}(\omega) \right)\right].
		\end{aligned}
	\label{eq:sdd}
\end{equation}
Comparing the second line of \cref{eq:sdd} to \cref{eq:ideal readout spectrum}, we notice that there are extra terms present, namely those with at least one $\lambda$ out front.
They are due to additional terms containing $X_\delta$ and $P_\delta$ on the RHS of the equation of motion for $Q$.
In \cref{fig:quad_diag}, they are denoted by the two gray arrows pointing to $Q$.
Thus, this is an \emph{imperfection} of the BAE measurement.

In fact, the additional contributions stem from the spectrum of the \emph{counterrotating quadrature} $X_{-\delta}$ (i.e., with frequency $-\delta$) and correlations of that quadrature with $X_{\delta}$.
That prompts a remarkably literal interpretation of ``counterrotating terms''.
It also clarifies the origin of the measurement BA. The CR quadrature can be written
\begin{equation}
  X_{-\delta}(t)=\cos(2\delta t)X_{\delta}(t)+\sin(2\delta t)P_\delta(t),
  \label{eq:cr quadrature}
\end{equation}
which makes it apparent that the measurement picks up some information about $P_\delta$, the quadrature conjugate to $X_\delta$.

In contrast, the \emph{measurement backaction}, as discussed in \cref{sec:mechanical spectrum},
arises from the gray arrows emanating from $I$.
This correction is contained within $S_{X_\delta X_\delta}$ in \cref{eq:sdd}.

In order to distinguish the two contributions (the imperfection and BA), we plot three functions in \cref{fig:spectra}.
The full optical spectrum \eqref{eq:sdd} in blue (solid) encompasses both contributions. A perfect BAE measurement of the modified mechanical spectrum \cref{eq:ideal readout spectrum,eq:CR correction} is shown in yellow (dashed),
it picks up the thermal contribution and the BA.
Finally, the expected result in RWA is red (dotted), which corresponds to a perfect BAE measurement of an otherwise undriven mechanical oscillator.
The sidepeaks of the modified mechanical spectrum do not show in the yellow curve, as they are suppressed by $|\chi_c(\omega)|^2$.

\subsection{Mechanical and optical variances}
An important application of BAE measurements is determining the quadrature variance, necessary for the verification of quantum squeezing.
The BA imperils this by increasing the variance to
\begin{equation}
	\ev{X_\delta^2}=\ev{X_{\text{RWA}}^2}+2n_{\text{CR}},
	\label{eq:increased variance}
\end{equation}
where $\ev{X_{\text{RWA}}^2}=\ev{X_{\text{unmeasured}}^2}$
is the result calculated without CR terms and  (cf.~\eqref{eq:CR correction})
\begin{equation}
	2n_{\text{CR}}\equiv\int\frac{\dd\omega}{2\pi}S_{X_\delta X_\delta}^{\text{CR}}(\omega)
	=\frac{8G^2(\kappa/\Gamma_{\text{eff}}+1)}{(\Gamma_{\text{eff}}+\kappa)^2+16\omega_{m,\text{eff}}^2}.
	\label{eq:ncr}
\end{equation}
Perhaps the most surprising aspect about this result is that $n_{\text{CR}}$ is independent of the temperature of the mechanical bath, whether it is squeezed or not, and what quadrature we measure.
This fact is already realized on the level of the spectrum correction \cref{eq:CR correction}.
The reason, as we have noted above, is that the optical quadrature $I$ in the measurement cavity is independent of the mechanics and that the BA is solely due to this quadrature (see \cref{fig:quad_diag}).
Note that although we have written the change in terms of a number of phonons $2n_{\text{CR}}$, it is not thermal heating that causes this effect, but rather the extraction of information about the conjugate quadrature.

In the optical spectrum, we saw that sidepeaks appear (cf.\ \cref{fig:spectra}).
To get an approximation to the variance of the measured mechanical quadrature, we integrate over the main peak of \cref{eq:sdd}.
Here, we calculate the error of this method,
which can be used for underdamped oscillators $\Gamma_{\text{eff}}\ll \omega_{m,\text{eff}}$.
The weight of the main peak is
\begin{equation}
	\ev{X_{\text{meas}}^2}=\ev{X_{\text{RWA}}^2}
	+2n_{\text{CR}}\frac{16\omega_{m,\text{eff}}^2(2\Gamma_{\text{eff}}+\kappa)}{(\Gamma_{\text{eff}}+\kappa)(\Gamma_{\text{eff}}^2+16\omega_{m,\text{eff}}^2)},
	\label{eq:main peak}
\end{equation}
with $n_{\text{CR}}$ given by \cref{eq:ncr}. It can be experimentally accessed by integrating $S_{d_{\text{out}}\dagg d_{\text{out}}}^{(0)}(\omega)/(\kappa G^2|\chi_c(\omega)|^2)$ from $-\omega_{m,\text{eff}}$ to $\omega_{m,\text{eff}}$, see \cref{app:main peak}.
The second term is the BA.
For example, if $\Gamma_{\text{eff}}/\kappa=10^{-4}$, $G/\kappa=0.05$, and $\omega_m/\kappa=5$ (i.e., $\cc=10^6$),
the variance increases by $0.5$ (in units of the zero-point fluctuations of the mechanical oscillator). Since $n_{\text{CR}}\propto G^2$, stronger driving quickly makes this effect noticeable.
In typical experimental regimes, where $\Gamma_{\text{eff}}\ll(\omega_m,\kappa)$, the BA in \cref{eq:main peak} is well approximated by $2n_{\text{CR}}$,
such that $\ev{X_{\text{meas}}^2}\approx\ev{X_\delta^{2}}$ \eqref{eq:increased variance}, as desired.

Finally, we would like to gain some insights about the good and bad cavity limits of \cref{eq:ncr,eq:main peak}.
To give some intuition what these limits imply for the optical spectrum, consider the solution of \cref{eq:langevin1}
\begin{equation}
	d(t)=\int_{-\infty}^t\dd{\tau}e^{\frac{\kappa(\tau-t)}{2}}\left[ 2iG\cos(\omega_m\tau) x(\tau)+d_{\text{in}}(\tau)\right].
  \label{eq:d solution}
\end{equation}
In the good cavity limit ($\omega_m\gg\kappa$), the exponential decay is negligible over a period $2\pi/\omega_m$. 
Thus 
\begin{multline}
  	\int_0^{\frac{2\pi}{\omega_m}}\dd{\tau}\cos(\omega_m\tau)[\cos(\omega_m\tau)X_\delta(\tau)+\sin(\omega_m\tau)P_\delta(\tau)]\\
  	\approx (1/2)\int_0^{\frac{2\pi}{\omega_m}}\dd{\tau}X_\delta(\tau),
\end{multline}
since $X_\delta(\omega)$ and $P_\delta(\omega)$ are centered around $\omega=0$.
On the other hand, in the bad cavity limit $\kappa>\omega_m$ the photons leave the cavity faster than the change in coupling parameter, such that no averaging takes place.
A simple argument then shows that the contribution from $X_\delta$ is roughly three times as big as the contribution from $P_\delta$.

These properties are reflected in the variances \cref{eq:ncr,eq:main peak}.
In the bad cavity limit, where $\kappa\gg\omega_{m,\text{eff}}$, the number of added phonons is
\begin{equation}
	2n_{\text{CR}}\to \frac{8G^2}{\Gamma_{\text{eff}}(\Gamma_{\text{eff}}+\kappa)}.
	\label{eq:bad cavity ncr}
\end{equation}
In this regime the separation of the two drives is small compared to the bandwidth of the cavity,
and therefore the BA becomes significant ($n_{\text{CR}}\sim1$) at a cooperativity $\cc\sim1$, where $\cc\equiv 4G^2/\kappa\Gamma_{\text{eff}}$.
As we have seen, the reason is that resonant and CR terms couple with equal strength.

In contrast, the good cavity limit is $\kappa\gg\Gamma_{\text{eff}}$ and $\omega_{m,\text{eff}}\gg\kappa$, where \cref{eq:ncr} reduces to
\begin{equation}
	2n_{\text{CR}}\to\frac{\kappa G^2}{2\Gamma_{\text{eff}}\omega_{m,\text{eff}}^2}.
	\label{eq:good cavity ncr}
\end{equation}
This agrees with the perturbative result in Ref.~\cite{Clerk2008}.
\Cref{eq:good cavity ncr} tells us that the BA depends inversely on the effective mechanical dissipation rate.
Physically, this is because $\Gamma_{\text{eff}}$, the rate at which the mechanical oscillator relaxes to its steady state,
competes with the BA rate due to CR terms, which is independent of $\Gamma_{\text{eff}}$.
The BA also decreases with increasing $\omega_m$, as CR terms become less resonant.
We find that the BA becomes significant ($n_{\text{CR}}\sim1$) for a cooperativity $\cc\sim16\omega_m^2/\kappa^2$.
Therefore, in the good cavity regime, the CR terms are suppressed by a factor ${\sim}16\omega_m^2/\kappa^2$ in comparison to resonant terms, due to the mechanism described above.

\section{Generalization}\label{sec:general recipe}
Above, we have taken $H_{\text{rest}}=0$ for simplicity, but we can in fact take any $H_{\text{rest}}$ and write down the solution,
as long as we know the steady state of $H_{\text{OM}}+H_{\text{rest}}+H_{\text{baths}}$, i.e., without the BAE measurement ($G=0$).
Furthermore, we need
\begin{equation}
	[H_{\text{rest}}, I]=0.
	\label{eq:recipe condition}
\end{equation}
If the unmeasured solution is also periodic, the consideration in~\cite{Malz2016} applies: For incommensurate periods, only the stationary part will be picked up by the measurement, but if they are commensurate, some other Fourier components may enter the output spectrum.

To formulate a general theory, we collect all system (input) operators into a vector $\vec F$ ($\vec F_{\text{in}}$).
The Langevin equations read
\begin{equation}
	\dot{\vec F}(t)=\matr A(t)\vec F(t)+\matr L\vec F_{\text{in}}(t).
\end{equation}
For a time-independent $\matr A(t)=\matr A$, a Fourier transform yields
\begin{equation}
	\vec F(\omega)=\matr\chi(\omega)\matr L\vec F_{\text{in}}(\omega),
\end{equation}
with susceptibility matrix $\matr{\chi}(\omega)=(-i\omega-\matr A)^{-1}$.
We can do the same for a time-periodic $\matr A(t)=\matr A(t+T)$ once we have reformulated the problem in terms of Fourier components.
Then
\begin{equation}
	\vec F^{(n)}(\omega)=\sum_{m=-\infty}^\infty \matr{\chi}^{(n-m)}(\omega-n\delta)\matr L\vec F^{(m)}_{\text{in}}(\omega).
	\label{eq:general solution}
\end{equation}
If we know how to solve the unmeasured system ($G=0$), we know how to find $\matr\chi^{(n)}$ in that case. 
Importantly, \cref{eq:general solution} does not make any assumption about the noise $\vec F_{\text{in}}^{(n)}$,
apart from stationarity.
For this reason, the replacement in \cref{eq:replacement rule} above will leave the susceptibility matrix $\matr{\chi}^{(n)}(\omega)$ unchanged
and therefore can be used to calculate the measurement BA for all systems with suitable $H_{\text{rest}}$.
In the end, we can write down the scattering matrix in terms of the susceptibility matrix
\begin{equation}
  \matr S^{(n)}(\omega)=\delta_{0,n}\matr{1}-\matr L\matr\chi^{(n)}(\omega)\matr L,
  \label{eq:scattering matrix}
\end{equation}
which can be use to calculate the output fields
\begin{equation}
  \vec F_{\text{out}}^{(n)}(\omega)=\sum_m\matr S^{(n-m)}(\omega-n\delta)\vec F_{\text{in}}^{(m)}(\omega).
  \label{eq:fourier input-output}
\end{equation}

For examples how $\matr\chi$ can look like, see the detailed calculations in \cref{app:weak coupling,app:dual-beam squeezing,app:parametric squeezing,app:two-mode}.
\Cref{app:dual-beam squeezing} contains the setup used to produce bichromatic squeezing~\cite{Mari2009,Kronwald2013,Lecocq2015,Pirkkalainen2015,Wollman2015,Malz2016,Lei2016},
\cref{app:weak coupling} discusses a weak-coupling version of the former, which is easily adapted to other weakly coupled systems, and \cref{app:parametric squeezing} contains the parametric squeezing case,
which is an example with more than one independent, relevant component in $\matr\chi$.
Last is \cref{app:two-mode}, which extends the method here to two mechanical modes,
and covers a recently developed two-mode BAE measurement~\cite{Woolley2013,Ockeloen-Korppi2016}.

\section{Conclusion}\label{sec:conclusion}
Using the framework from Ref.~\cite{Malz2016}, we derived the full solution of an optomechanical system subject to a dual-beam BAE measurement~\cite{Bocko1996,Hertzberg2010,Suh2014,Lecocq2015,Lei2016}.
This enables us to calculate the modification of the spectrum and quantify the measurement backaction precisely.
Furthermore, we demonstrate that our technique is versatile, by showing how to generalize the calculation to systems where the mechanical resonator is additionally coupled to other degrees of freedom and illustrate the technique with several examples. 

\section*{Acknowledgments} 
We are grateful to Chan Lei and John Teufel for stimulating and insightful discussions. A.N. holds a University Research Fellowship from the Royal Society and acknowledges additional support from the Winton Programme for the Physics of Sustainability. D.M. acknowledges support by the UK Engineering and Physical Sciences Research Council (EPSRC) under Grant No. EP/M506485/1.

\appendix
\section{Dissipative bichromatic squeezing}\label{app:dual-beam squeezing}
An important application of the type of BAE measurement discussed in this article is the verification of quantum squeezing in mechanical resonators, e.g., Refs.~\cite{Lecocq2015,Lei2016}. Here we generalize our method to this squeezing scheme,
proposed in \cite{Mari2009,Kronwald2013}, which has successfully produced quantum squeezing of the resonator~\cite{Wollman2015,Pirkkalainen2015,Lecocq2015,Lei2016}.
In this case
\begin{equation}
	H_{\text{rest}}=-\Delta c\dagg c-\left[ c\left( G_+e^{2i\delta t} b +G_-b\dagg\right)+\text{h.c.} \right],
\end{equation}
where we have already displaced and linearized.
$c$ is the annihilation operator of another cavity in a frame rotating with the lower-frequency drive.
Furthermore, we have applied the RWA to $H_{\text{rest}}$.
The governing Langevin equations are
\begin{subequations}
	\begin{align}
		\dot c&=\left( i\Delta-\frac{\kappa_2}{2} \right)c+\sqrt{\kappa_2}c_{\text{in}}+i\left(G_-b+G_+e^{-2i\delta t}b\dagg\right),\\
		\dot b&=\left( -i\omega_m-\frac{\gamma}{2} \right)b+\sqrt{\gamma}b_{\text{in}}+i\left( G_-c+G_+e^{-2i\delta t}c\dagg \right),
	\end{align}	\label{eq:dissipative squeezing langevin}
\end{subequations}
where $c_{\text{in}}$ corresponds to a vacuum (zero temperature) bath, and $b_{\text{in}}$ to a finite temperature bath with mean occupation $n_{\text{th}}$,
so $\ev{c_{\text{in}}(t)c_{\text{in}}(t')\dagg}=\delta(t-t')$,
$\ev{b_{\text{in}}(t)\dagg b_{\text{in}}(t')}=\delta(t-t')n_{\text{th}}$,
$\ev{b_{\text{in}}(t)b_{\text{in}}(t')\dagg}=\delta(t-t')(1+n_{\text{th}})$ (other correlators are zero).

In the Floquet ansatz, \cref{eq:dissipative squeezing langevin} can be expressed in terms of a block-diagonal infinite matrix~\cite{Malz2016}
with blocks
\begin{widetext}
\begin{equation}
  \label{eq:decoupled blocks}
  \mat{\chi_{\text{opt}}^{-1}(\omega-n\delta)&-iG_-&0&-iG_+\\
  -iG_-&\chi_m^{-1}(\omega-n\delta)&-iG_+&0\\
  0&iG_+&\chi_{\text{opt}}^{-1*}(-\omega+(n+2)\delta)&iG_-\\
  iG_+&0&iG_-&\chi_m^{-1*}(-\omega+(n+2)\delta)}
  \mat{c^{(n)}(\omega)\\b^{(n)}(\omega)\\c^{(n+2)\dag}(\omega)\\b^{(n+2)\dag}(\omega)}
  =\mat{\sqrt{\kappa}c_{\text{in}}^{(n)}(\omega)\\\sqrt{\gamma}b_{\text{in}}^{(n)}(\omega)\\
  \sqrt{\kappa}c_{\text{in}}^{(n+2)\dag}(\omega)\\ \sqrt{\gamma}b_{\text{in}}^{(n+2)\dag}(\omega)},
\end{equation}
\end{widetext}
where the cavity and mechanical response functions read $\chi_{\text{opt}}(\omega)=[\kappa/2-i(\omega+\Delta)]^{-1}\neq\chi_c$ and $\chi_m(\omega)=[\gamma/2-i(\omega-\omega_m)]^{-1}$, respectively.

Blocks with zero input will vanish in the steady state.
Since the noise resides entirely in the zeroth Fourier component (it describes a stationary process),
\begin{equation}
	c_{\text{in}}^{(n)}=b_{\text{in}}^{(n)}=0,\qquad\forall n\neq0.
\end{equation}

Although the result for general $\delta$ and $\Delta$ is available~\cite{Malz2016}, we focus on the simple and relevant case $\delta=-\Delta=\omega_m$, where
\begin{equation}
	\mat{b^{(0)}(\omega)\\ b^{(2)\dag}(\omega)}
	=J(\omega)\mat{1&iG_-\chi_{\text{opt}}(\omega)\\0&iG_+\chi_{\text{opt}}(\omega)}\mat{\sqrt{\gamma}b_{\text{in}}\\ \sqrt{\kappa_2}c_{\text{in}}},
	\label{eq:dual-beam unmeasured}
\end{equation}
with 
$	J(\omega)=\left[ \chi_m^{-1}(\omega)+\G^2\chi_{\text{opt}}(\omega) \right]^{-1}.$

Now we include the noise of a BAE measurement as per \cref{sec:general recipe}.
The measurement requires another cavity mode coupled to the mechanical oscillator.
We define
\begin{equation}
	\vec F=\mat{b&b\dagg&c&c\dagg},
	\quad \vec F_{\text{in}}=\mat{b_{\text{in}}&b_{\text{in}}\dagg&c_{\text{in}}&c_{\text{in}}\dagg}.
	\label{eq:system vector}
\end{equation}

From \cref{eq:dual-beam unmeasured}, we can directly infer the elements
\begin{equation}
	\mat{\matr\chi_{11}^{(0)}(\omega)&\matr\chi_{13}^{(0)}(\omega)\\
	\matr\chi_{21}^{(2)}(\omega-2\delta)&\matr\chi_{23}^{(2)}(\omega-2\delta)}
	=J(\omega)\mat{1&iG_-\chi_{\text{opt}}(\omega)\\0&iG_+\chi_{\text{opt}}(\omega)}.
\end{equation}
Furthermore, the elements of $\matr\chi$ are not all independent. In general, 
\begin{equation}
	\chi_{ij}^{(n)*}(\omega)=\chi_{\bar i\bar j}^{(-n)}(-\omega),
	\label{eq:chi symmetry}
\end{equation}
where $\bar i$ is the index for the operator that is the hermitian conjugate of the operator indexed by $i$ (so $\bar 1=2$, $\bar 3=4$, and \emph{vice versa}).

We are only interested in the elements that connect the system operators with the mechanical bath, as those will determine the susceptibility to the measurement noise. 
Fortunately, the only non-zero ones are
\begin{equation}
	\label{eq:full squeezing chi}
	\matr\chi_{11}^{(0)}(\omega)=J(\omega)=\matr\chi_{22}^{(0)*}(-\omega).
\end{equation}
This gives the problem exactly the same structure as the case with $H_{\text{rest}}=0$, except that here $\chi_m(\omega)$ is replaced by $J(\omega)$.

Nevertheless, we give the general formula for the added contribution due to the measurement
\begin{subequations}
  \begin{align}
  	  b_{\text{add}}^{(n)}(\omega)&=2\sqrt{\kappa}G\sum_mK^{(n-m)}(\omega-n\delta)f_{\text{in}}^{(m)}(\omega),\\
  	  b_{\text{add}}^{(n)\dag}(\omega)&=2\sqrt{\kappa}G\sum_mK^{(-n+m)*}(-\omega+n\delta)f_{\text{in}}^{(m)}(\omega),
  \end{align}
  \label{eq:general added noise}
\end{subequations}
where
\begin{equation}
	K^{(n)}(\omega)\equiv \matr\chi_{11}^{(n)}(\omega)-\matr\chi_{12}^{(n)}(\omega).
\end{equation}
This gives (note similarity with \cref{eq:mech fourier cpts})
\begin{equation}
	b^{(n)}_{\text{add}}=2\sqrt{\kappa}G J(\omega-n\delta)f_{\text{in}}^{(n)}(\omega).
\end{equation}
and hermitian conjugate.
The CR correction can be calculated as before and looks very familiar (cf.\ \cref{eq:CR correction})
\begin{multline}
	S_{X_{\delta}X_{\delta}}^{\text{CR}}(\omega)=\kappa\lambda^2 G^2\left( |J(\omega+\delta)\chi_c(\omega+2\delta)|^2\right.\\
	\left.+|J(-\omega+\delta)\chi_c(-\omega+2\delta)|^2 \right).
	\label{eq:dissipative CR}
\end{multline}
The reason why it is so simple is that here $b^{(0)}(\omega)$ only couples to $b_{\text{in}}$ and not to $b_{\text{in}}\dagg$.
This ceases to be the case for $\Delta\neq-\omega_m$, or in parametric squeezing in \cref{app:parametric squeezing}.

\section{Dissipative bichromatic squeezing in the absence of strong-coupling effects}\label{app:weak coupling}
As a relevant example for how a weakly coupled $H_{\text{rest}}$ can lead to effective parameters in the mechanical susceptibility, we consider the weak-coupling version of \cref{app:dual-beam squeezing}.
We place the red-detuned drive on the red sideband $\Delta=-\omega_m$, but allow the other to vary.
To second order in $G_\pm$, the effective quantum Langevin equation is (in a frame rotating with frequency $\delta$)
\begin{equation}
	\dot b=-\left[i(\omega_{m,\text{eff}}-\delta)+\frac{\Gamma_{\text{eff}}}{2} \right]b+\sqrt{\gamma}b_{\text{in}}+\frac{2i\G}{\sqrt{\kappa}}s_{\text{in}},
	\label{eq:weak QLE}
\end{equation}
with~\cite{Malz2016}
\begin{subequations}
	\begin{align}
		\Gamma_{\text{eff}}&=\gamma+\frac{4}{\kappa}\left( G_-^2-\frac{G_+^2}{1+4\eps^2/\kappa^2} \right),\\
		\omega_{m,\text{eff}}&=\omega_m+\frac{G_+^2\eps}{(\kappa/2)^2+\eps^2},\\
		s_{\text{in}}&=\frac{G_-d_{\text{in}}+G_+d_{\text{in}}\dagg}{\G}.
	\end{align}	\label{eq:effective parameters}
\end{subequations}
Here, $\eps\equiv2(\delta-\omega_m)$ and $\G\equiv\sqrt{G_-^2-G_+^2}$.
$s_{\text{in}}$ is a Bogoliubov rotation of the original optical bath operators, and is therefore a squeezed, vacuum bath with nonzero anomalous correlators, such as $\ev{s_{\text{in}}(\omega)s_{\text{in}}(\omega')}=-2\pi\delta(\omega+\omega')G_+G_-/\G^2$.

The equivalent master equation (generalization of Ref.~\cite{Kronwald2013} to general drive detuning $\delta$) is
\begin{multline}
	\dot\dens=-i[(\omega_{m,\text{eff}}-\delta)b\dagg b,\dens]+\left\{\Gamma_{\text{eff}}(n_{\text{eff}}+1)\D[b]\right.\\\left.
	+\Gamma_{\text{eff}} n_{\text{eff}}\D[b\dagg]+\frac{4\G^2}{\kappa}\D[\beta]\right\}\dens,
	\label{eq:weak QME}
\end{multline}
where $\D[a]\dens\equiv a\dens a\dagg-\frac{1}{2}(a\dagg a\dens+\dens a\dagg a)$ is the Lindblad superoperator.

In order to include the measurement noise, it is easier to work in the laboratory frame, where 
\begin{equation}
	\dot b=-\left(i\omega_{m,\text{eff}}+\frac{\Gamma_{\text{eff}}}{2}\right)b+\sqrt{\gamma}b_{\text{in}}
	+2\sqrt{\kappa}G f_{\text{in}}+\frac{2i\G}{\sqrt{\kappa}}s_{\text{in}}.
	\label{eq:measured weak QLE}
\end{equation}
There is a subtlety here, as the anomalous averages of $s_{\text{in}}$, will become rotating in the transition from a rotating frame into the laboratory frame.
We can ignore this difficulty, because we are only interested in the correction.
Rewriting \cref{eq:measured weak QLE} in Fourier components, we find that the only independent nonzero element of the susceptibility matrix is 
\begin{equation}
	\matr\chi_{11}^{(0)}(\omega)=\chi_{m,\text{eff}}(\omega)\equiv[\Gamma_{\text{eff}}/2-i(\omega-\omega_{m,\text{eff}})]^{-1}.
\end{equation}
We could have obtained this from \cref{eq:full squeezing chi} by setting $\G=0$ (weak coupling) and replacing $\omega_m$ and $\gamma$ by their modified values.
Thus we can use the result \cref{eq:dissipative CR} with $J\to\chi_{m,\text{eff}}$.
Therefore, the results in the article are valid in the weak coupling case as well, with $\gamma\to\Gamma_{\text{eff}}$ and $\omega_m\to\omega_{m,\text{eff}}$.

Optionally, we can remain in a rotating frame, but that means we have to rotate the added measurement noise to $\tilde f_{\text{in}}\equiv e^{i\delta t}f_{\text{in}}$.
This implies $\tilde f_{\text{in}}^{(n)}=f_{\text{in}}^{(n-1)}$.
Solving the Langevin equation gives $\chi_{11}^{(0)}(\omega)=[\Gamma_{\text{eff}}/2-i(\omega-\omega_{m,\text{eff}}+\delta)]^{-1}$.
Consulting \cref{eq:general added noise}, we find that the new set of Fourier components $\tilde b_{\text{add}}^{(n)}(\omega)=b_{\text{add}}^{(n-1)}(\omega)$.
It is straightforward to check that in the end $\tilde b_{\text{add}}(t)=e^{i\delta t}b_{\text{add}}(t)$.

This result can be adapted to a wide variety of cases, as long as they can be approximated by coupling the harmonic oscillators to baths only (and potentially modify its effective parameters).

\section{Parametric squeezing}\label{app:parametric squeezing}
Squeezing is induced naturally in a degenerate parametric amplifier~\cite{Walls2008}.
Whilst it is limited to 3\,dB of squeezing, its solution has other interesting aspects. Here,
\begin{equation}
	H_{\text{rest}}=\omega_mb\dagg b+(\mu b^2e^{2i\omega_m t}+\text{h.c.}),
\end{equation}
where $\mu$ is the parametric driving strength.
Without the BAE measurement $(G=0$), the quantum Langevin equation is
\begin{equation}
	\dot b=-i\omega_m b-i\mu b\dagg e^{-2i\omega_m t}-\frac{\gamma}{2}b+\sqrt{\gamma}b_{\text{in}}.
\end{equation}
In Fourier components,
\begin{multline}
	\mat{\chi_m^{-1}(\omega-n\omega_m)&i\mu\\-i\mu&\chi_m^{-1}(\omega-n\omega_m)}\mat{b^{(n)}(\omega)\\b^{(n+2)\dag}(\omega)}\\
	=\sqrt{\gamma}\mat{\delta_{n,0}b_{\text{in}}\\\delta_{n,2}b_{\text{in}}\dagg},
\end{multline}
where $\chi_m(\omega)\equiv\left[ \gamma/2-i(\omega-\omega_m) \right]^{-1}$.
Inverting leads to
\begin{multline}
\mat{b^{(n)}(\omega)\\b^{(n+2)\dag}(\omega)}=A(\omega-n\omega_m)\\\times
	\mat{\chi_m^{-1}(\omega-n\omega_m)&-i\mu\\i\mu&\chi_m^{-1}(\omega-n\omega_m)}
	\mat{\sqrt{\gamma}\delta_{n,0}b_{\text{in}}\\ \sqrt{\gamma}\delta_{n,2}b_{\text{in}}\dagg},
\end{multline}
where
\begin{equation}
	A(\omega)\equiv\left[ \chi_m^{-2}(\omega-n\omega_m)-\mu^2 \right]^{-1}.
	\label{eq:A}
\end{equation}
It is again enough to consider only one block (the other nonzero block is the hermitian conjugate)
\begin{equation}
	\mat{b^{(0)}(\omega)\\b^{(2)\dag}(\omega)}=A(\omega)
	\mat{\chi_m^{-1}(\omega)&-i\mu\\i\mu&\chi_m^{-1}(\omega)}
	\mat{\sqrt{\gamma}b_{\text{in}}\\0}.
\end{equation}

To make the discussion as simple as possible, we choose $\delta=\omega_m$ for the measurement scheme.
Then the same analysis as above can be applied to obtain
\begin{subequations}
	\begin{align}
		\matr\chi^{(0)}_{11}(\omega)&=A(\omega)\chi_m^{-1}(\omega),\\
		\matr\chi^{(2)}_{12}(\omega-2\omega_m)&=-i\mu A(\omega),\\
		\matr\chi^{(-2)*}_{12}(-\omega+2\omega_m)&=i\mu A^*(\omega).
	\end{align}
\end{subequations}

In order to evaluate the added noise due to the BAE measurement, we have to add the noise~\cref{eq:time-dependent bath}
and use \cref{eq:general added noise}.
We find ($\lambda$ labels terms with CR origin)
\begin{equation}
	\begin{aligned}
	b_{\text{add}}^{(n)}&(\omega)=i\sqrt{\kappa}G\chi_c(\omega)I_{\text{in}}(\omega)\\
	&\times
	\begin{cases}
		i\mu A(\omega-\delta)&\text{if }n=3\\
		i\mu \lambda A(\omega+\delta)+A(\omega-\delta)\chi_m^{-1}(\omega-\delta)&\text{if }n=1\\
		\lambda A(\omega+\delta)\chi_m^{-1}(\omega+\delta)&\text{if }n=-1
	\end{cases}
	\end{aligned}
	\label{eq:parametric badd}
\end{equation}
This reverts to \cref{eq:mech fourier cpts} when we set $\mu=0$.
Using \cref{eq:rotating spectrum} with the spectra calculated from \cref{eq:parametric badd}, we obtain the correction to the quadrature spectrum.
The resulting spectra have terms rotating at multiples of $\delta$. They could be measured by coupling to another suitably driven cavity mode~\cite{Malz2016}, but tend to be very small.

\section{Two-mode BAE}\label{app:two-mode}
In this section, we consider another recently experimentally demonstrated QND scheme~\cite{Woolley2013,Ockeloen-Korppi2016}.
The goal here is to measure a collective quadrature of two mechanical oscillators, in order to eventually measure both quadratures of an external force with the same device.
Whilst the overall theory is more general \cite{Tsang2012}, in the specific experiment we consider here,
both mechanical resonators couple to the same cavity and the problem has the Hamiltonian~\cite{Woolley2013}
\begin{equation}
  \begin{aligned}
  	H&=(\omega_m+\Omega)b_1\dagg b_1+(\omega_m-\Omega)b_2\dagg b_2+\omega_{\text{cav}}a\dagg a\\
  	&+g_1(b_1+b_1\dagg)a\dagg a	+g_2(b_2+b_2\dagg)a\dagg a+H_{\text{drive}}+H_{\text{diss}}
  \end{aligned}
  \label{eq:two-mode hamiltonian}
\end{equation}
Note that in Ref.~\cite{Woolley2013} the mechanical oscillators have annihilation operators $a$ and $b$, and the cavity $c$,
whereas here, they are $b_1,b_2$, and $a$, respectively.
As before, bichromatic driving leads to a coherent state
\begin{equation}
  	\hat a=2\cos(\Omega t)\bar a+\hat d,
\end{equation}
where $\bar a$ is a real number, and $\langle\hat d\rangle=0$.

Our generic solution is applicable, because \cref{eq:recipe condition} is fulfilled after linearizing,
no matter which of the oscillators we put into $H_{\text{rest}}$. For example, we could chose
\begin{equation}
	H_{\text{rest,1}}=(\omega_m+\Omega)b_1\dagg b_1+g_1(b_1+b_1\dagg)2\cos(\Omega t)\bar a(d+d\dagg),
\end{equation}
but the choice with $1\to2$ is equally valid.
Our result for the backaction on the mechanical oscillators (all of \cref{sec:framework}, particularly \cref{eq:CR correction}) thus applies to both resonators individually.

The correction to the cavity spectrum is slightly more tricky to find, since the fluctuations in the two resonators are correlated, leading to cross terms. 
Instead of \cref{eq:langevin1} we have
\begin{equation}
  \dot d=-\frac{\kappa}{2}d+\sqrt{\kappa}d_{\text{in}}+2i\cos(\Omega t)
  \left(G_1x_1+G_2x_2\right),
  \label{eq:d eom two-mode}
\end{equation}
where $x_{1,2}$ are the position operators for the two oscillators, and $G_{1,2}\equiv g_{1,2}\bar a$.
$x_{1,2}$ are both given through \cref{eq:mech fourier cpts}, but with the respective parameters for each resonator.

To calculate the optical spectrum, first note that there is a cross correlation.
For $m,n\in\{1,-1\}$, we have
\begin{equation}
  \begin{aligned}
  	&\int\frac{\dd{\omega'}}{2\pi}\ev{x_1^{(m)}(\omega+m\delta)x_2^{(n)}(\omega')}
  	=-\kappa G_1G_2\\
  	&\times|\chi_c(\omega+m\delta)|^2\chi_{x_1}(\omega)\chi_{x_2}(-\omega-(m+n)\delta),
  \end{aligned}
  \label{eq:cross term}
\end{equation}
with $\chi_{x_{1,2}}$ defined analogously to $\chi_x$ in \cref{app:optical spectrum}.
Then $S_{d\dagg d}^{(0)}$ contains two copies of \cref{eq:sdd} (one for each resonator), in addition to
\begin{equation}
  \begin{aligned}
  	&4G^2|\chi_c(\omega)|^2\Re\left[ S_{x_1x_2}^{(0)}(\omega-\delta) + S_{x_1x_2}^{(0)}(\omega+\delta)\right.\\
   	 &\qquad+\left.S_{x_1x_2}^{(2)}(\omega-\delta)+ S_{x_1x_2}^{(-2)}(\omega+\delta)\right],
  \end{aligned}
\end{equation}
where $S_{x_1x_2}$ are given through \cref{eq:fourier components,eq:cross term}.
Here, we used the property $[S_{AB}^{(n)}(\omega)]\dagg=S_{B\dagg A\dagg}^{(-n)}(\omega+n\delta)$~\cite{Malz2016}.

This demonstrates that our technique can also be applied for multiple modes coupled to the cavity, as long as $[H_{\text{rest}},I]=0$.

\section{Integration of the main peak}\label{app:main peak}
One goal of a BAE measurement of a mechanical oscillator quadrature is to extract the quadrature variance.
We show that the weight of the main peak of the optical spectrum
\begin{equation}
	\begin{aligned}
		&\frac{S_{d\dagg d}^{(0)}(\omega)}{G^2|\chi_c(\omega)|^2}=S_{xx}^{(0)}(\omega-\delta)+S_{xx}^{(0)}(\omega+\delta)\\
		&\qquad+S_{xx}^{(2)}(\omega-\delta)+S_{xx}^{(-2)}(\omega+\delta)
	\end{aligned}
\end{equation}
is a good measure for the quadrature variance. 
The weight of its middle peak at $\omega=0$ is
\begin{equation}
	W=\int_{-\infty}^{\infty}\frac{\dd{\omega}}{2\pi}S_{xx}^{(0)}(\omega)+2\Re[S_{xx}^{(2)}(\omega)].
	\label{eq:main peak weight formula}
\end{equation}
\Cref{eq:main peak weight formula} can be evaluated with the formulae in \cref{app:optical spectrum}.
Given the output spectrum $\kappa S_{d\dagg d}^{(0)}(\omega)$,
the weight can be approximated by integrating from $-\omega_m$ to $\omega_m$
\begin{equation}
	W\approx\ev{X_{\text{meas}}^2}\equiv\int_{-\omega_m}^{\omega_m}\frac{\dd{\omega}}{2\pi}\frac{S_{d_{\text{out}}\dagg d_{\text{out}}}^{(0)}(\omega)}{\kappa G^2|\chi_c(\omega)|^2}.
	\label{eq:experimental main peak}
\end{equation}

\section{Full quadrature spectrum}\label{app:full quadrature spectrum}
Here we present the full expression for the spectrum of the rotating quadrature $X_\delta$
\begin{widetext}
\begin{equation}\label{eq:full quadrature spectrum}
	\begin{aligned}
		S_{X_\delta X_\delta}^{(0)}(\omega)&=
		\gamma\left(|\chi_m(\omega+\delta)|^2 (n_{\text{th}}+1)+|\chi_m(-\omega+\delta)|^2n_{\text{th}}\right)
		+\kappa G^2\left\{-2|\chi_c(\omega)|^2\Re\left[ \chi_m(\omega+\delta)\chi_m(-\omega+\delta) \right]\right.\\
		&+\left.|\chi_m(\omega+\delta)|^2\left( |\chi_c(\omega)|^2+|\chi_c(\omega+2\delta)|^2 \right)
		+|\chi_m(-\omega+\delta)|^2\left( |\chi_c(\omega-2\delta)|^2+|\chi_c(\omega)|^2 \right)\right\}.
	\end{aligned}
\end{equation}
It contains a part due to the thermal bath of the oscillator (terms with a $\gamma$) and an optical part (terms with $\kappa$).
If $\delta=\omega_m$, $\chi_m(-\omega+\delta)=\chi_m^*(\omega+\delta)$, and the negative term in the curly brackets cancels two resonant contributions in the second line.
This cancellation makes the measurement BAE in RWA. The two terms left over cause the two side peaks to appear.
Inverting the sign of the first term in curly brackets would lead to the spectrum of the conjugate quadrature $S_{P_\delta P_\delta}^{(0)}(\omega)$, which gets all the BA in RWA. The BA due to CR terms is the same in both.
If the oscillator is weakly coupled to other degrees of freedom, this formula is still approximately correct, using effective parameters $\Gamma_{\text{eff}}$, $\omega_{m,\text{eff}}$ and $n_{\text{eff}}$.

\section{Optical spectrum}\label{app:optical spectrum}
In this section we outline the derivation of the optical spectrum. We use the expression given in the main text \eqref{eq:sdd}.
For $H_{\text{rest}}=0$, the necessary correlators are
\begin{equation}
	\int\frac{\dd{\omega'}}{2\pi}\ev{x^{(0)}(\omega)x^{(0)}(\omega')}=\gamma\left[ |\chi_m(-\omega)|^2n_{\text{th}}
		+|\chi_m(\omega)|^2(n_{\text{th}}+1)\right],
\end{equation}
and for $m,n\in\{1,-1\}$
\begin{equation}
	\int\frac{\dd{\omega'}}{2\pi}\ev{x^{(m)}(\omega+m\delta)x^{(n)}(\omega')}
	=-\kappa G^2|\chi_c(\omega+m\delta)|^2\chi_x(\omega)\chi_x(-\omega-(m+n)\delta),
\end{equation}
where
$	\chi_x(\omega)\equiv\chi_m(\omega)-\lambda\chi_m^*(-\omega).$
This gives
\begin{subequations}
	\begin{align}
		S_{xx}^{(0)}(\omega)&=\gamma\left[ |\chi_m(-\omega)|^2n_{\text{th}}+|\chi_m(\omega)|^2(n_{\text{th}}+1) \right]
			+\kappa G^2\left( |\chi_x(-\omega)|^2|\chi_c(\omega+\delta)|^2+|\chi_x(\omega)|^2|\chi_c(\omega-\delta)|^2 \right),\\
		S_{xx}^{(2)}(\omega)&=-\kappa G^2|\chi_c(\omega+\delta)|^2\chi_x^*(-\omega)\chi_x^*(\omega+2\delta),\\
		S_{xx}^{(-2)}(\omega)&=-\kappa G^2|\chi_c(\omega-\delta)|^2\chi_x(\omega)\chi_x(-\omega+2\delta),
	\end{align}
\end{subequations}
whence $S_{d\dagg d}^{(0)}(\omega)$ can be constructed, using (cf.~\cref{eq:sdd})
\begin{equation}
	S_{d\dagg d}^{(n)}(\omega)=G^2\chi_c(\omega)\chi_c^*(\omega+n\delta)\left[ S_{xx}^{(n)}(\omega-\delta)+S_{xx}^{(n)}(\omega+\delta)
		+S_{xx}^{(n+2)}(\omega-\delta)+S_{xx}^{(n-2)}(\omega+\delta)\right].
\end{equation}
\end{widetext}

\bibliographystyle{apsrev4-1.bst}
\bibliography{library}{}

\begin{thebibliography}{23}%
\makeatletter
\providecommand \@ifxundefined [1]{%
 \@ifx{#1\undefined}
}%
\providecommand \@ifnum [1]{%
 \ifnum #1\expandafter \@firstoftwo
 \else \expandafter \@secondoftwo
 \fi
}%
\providecommand \@ifx [1]{%
 \ifx #1\expandafter \@firstoftwo
 \else \expandafter \@secondoftwo
 \fi
}%
\providecommand \natexlab [1]{#1}%
\providecommand \enquote  [1]{``#1''}%
\providecommand \bibnamefont  [1]{#1}%
\providecommand \bibfnamefont [1]{#1}%
\providecommand \citenamefont [1]{#1}%
\providecommand \href@noop [0]{\@secondoftwo}%
\providecommand \href [0]{\begingroup \@sanitize@url \@href}%
\providecommand \@href[1]{\@@startlink{#1}\@@href}%
\providecommand \@@href[1]{\endgroup#1\@@endlink}%
\providecommand \@sanitize@url [0]{\catcode `\\12\catcode `\$12\catcode
  `\&12\catcode `\#12\catcode `\^12\catcode `\_12\catcode `\%12\relax}%
\providecommand \@@startlink[1]{}%
\providecommand \@@endlink[0]{}%
\providecommand \url  [0]{\begingroup\@sanitize@url \@url }%
\providecommand \@url [1]{\endgroup\@href {#1}{\urlprefix }}%
\providecommand \urlprefix  [0]{URL }%
\providecommand \Eprint [0]{\href }%
\providecommand \doibase [0]{http://dx.doi.org/}%
\providecommand \selectlanguage [0]{\@gobble}%
\providecommand \bibinfo  [0]{\@secondoftwo}%
\providecommand \bibfield  [0]{\@secondoftwo}%
\providecommand \translation [1]{[#1]}%
\providecommand \BibitemOpen [0]{}%
\providecommand \bibitemStop [0]{}%
\providecommand \bibitemNoStop [0]{.\EOS\space}%
\providecommand \EOS [0]{\spacefactor3000\relax}%
\providecommand \BibitemShut  [1]{\csname bibitem#1\endcsname}%
\let\auto@bib@innerbib\@empty
\bibitem [{\citenamefont {Caves}\ \emph {et~al.}(1980)\citenamefont {Caves},
  \citenamefont {Thorne}, \citenamefont {Drever}, \citenamefont {Sandberg},\
  and\ \citenamefont {Zimmermann}}]{Caves1980}%
  \BibitemOpen
  \bibfield  {author} {\bibinfo {author} {\bibfnamefont {C.~M.}\ \bibnamefont
  {Caves}}, \bibinfo {author} {\bibfnamefont {K.~S.}\ \bibnamefont {Thorne}},
  \bibinfo {author} {\bibfnamefont {R.~W.~P.}\ \bibnamefont {Drever}}, \bibinfo
  {author} {\bibfnamefont {V.~D.}\ \bibnamefont {Sandberg}}, \ and\ \bibinfo
  {author} {\bibfnamefont {M.}~\bibnamefont {Zimmermann}},\ }\href {\doibase
  10.1103/RevModPhys.52.341} {\bibfield  {journal} {\bibinfo  {journal}
  {Reviews of Modern Physics}\ }\textbf {\bibinfo {volume} {52}},\ \bibinfo
  {pages} {341} (\bibinfo {year} {1980})}\BibitemShut {NoStop}%
\bibitem [{\citenamefont {Clerk}\ \emph {et~al.}(2010)\citenamefont {Clerk},
  \citenamefont {Devoret}, \citenamefont {Girvin}, \citenamefont {Marquardt},\
  and\ \citenamefont {Schoelkopf}}]{Clerk2010}%
  \BibitemOpen
  \bibfield  {author} {\bibinfo {author} {\bibfnamefont {A.~A.}\ \bibnamefont
  {Clerk}}, \bibinfo {author} {\bibfnamefont {M.~H.}\ \bibnamefont {Devoret}},
  \bibinfo {author} {\bibfnamefont {S.~M.}\ \bibnamefont {Girvin}}, \bibinfo
  {author} {\bibfnamefont {F.}~\bibnamefont {Marquardt}}, \ and\ \bibinfo
  {author} {\bibfnamefont {R.~J.}\ \bibnamefont {Schoelkopf}},\ }\href
  {\doibase 10.1103/RevModPhys.82.1155} {\bibfield  {journal} {\bibinfo
  {journal} {Reviews of Modern Physics}\ }\textbf {\bibinfo {volume} {82}},\
  \bibinfo {pages} {1155} (\bibinfo {year} {2010})}\BibitemShut {NoStop}%
\bibitem [{\citenamefont {Braginsky}\ \emph {et~al.}(1980)\citenamefont
  {Braginsky}, \citenamefont {Vorontsov},\ and\ \citenamefont
  {Thorne}}]{Braginsky1980}%
  \BibitemOpen
  \bibfield  {author} {\bibinfo {author} {\bibfnamefont {V.~B.}\ \bibnamefont
  {Braginsky}}, \bibinfo {author} {\bibfnamefont {Y.~I.}\ \bibnamefont
  {Vorontsov}}, \ and\ \bibinfo {author} {\bibfnamefont {K.~S.}\ \bibnamefont
  {Thorne}},\ }\href {\doibase 10.1126/science.209.4456.547} {\bibfield
  {journal} {\bibinfo  {journal} {Science}\ }\textbf {\bibinfo {volume}
  {209}},\ \bibinfo {pages} {547} (\bibinfo {year} {1980})}\BibitemShut
  {NoStop}%
\bibitem [{\citenamefont {Abbott}\ and\ \citenamefont
  {Others}(2016)}]{Abbott2016}%
  \BibitemOpen
  \bibfield  {author} {\bibinfo {author} {\bibfnamefont {B.~P.}\ \bibnamefont
  {Abbott}}\ and\ \bibinfo {author} {\bibnamefont {Others}},\ }\href {\doibase
  10.1103/PhysRevLett.116.061102} {\bibfield  {journal} {\bibinfo  {journal}
  {Physical Review Letters}\ }\textbf {\bibinfo {volume} {116}},\ \bibinfo
  {pages} {061102} (\bibinfo {year} {2016})},\ \Eprint
  {http://arxiv.org/abs/1602.03837} {arXiv:1602.03837} \BibitemShut {NoStop}%
\bibitem [{\citenamefont {Bocko}\ and\ \citenamefont
  {Onofrio}(1996)}]{Bocko1996}%
  \BibitemOpen
  \bibfield  {author} {\bibinfo {author} {\bibfnamefont {M.~F.}\ \bibnamefont
  {Bocko}}\ and\ \bibinfo {author} {\bibfnamefont {R.}~\bibnamefont
  {Onofrio}},\ }\href {\doibase 10.1103/RevModPhys.68.755} {\bibfield
  {journal} {\bibinfo  {journal} {Reviews of Modern Physics}\ }\textbf
  {\bibinfo {volume} {68}},\ \bibinfo {pages} {755} (\bibinfo {year}
  {1996})}\BibitemShut {NoStop}%
\bibitem [{\citenamefont {Clerk}\ \emph {et~al.}(2008)\citenamefont {Clerk},
  \citenamefont {Marquardt},\ and\ \citenamefont {Jacobs}}]{Clerk2008}%
  \BibitemOpen
  \bibfield  {author} {\bibinfo {author} {\bibfnamefont {A.~A.}\ \bibnamefont
  {Clerk}}, \bibinfo {author} {\bibfnamefont {F.}~\bibnamefont {Marquardt}}, \
  and\ \bibinfo {author} {\bibfnamefont {K.}~\bibnamefont {Jacobs}},\ }\href
  {\doibase 10.1088/1367-2630/10/9/095010} {\bibfield  {journal} {\bibinfo
  {journal} {New Journal of Physics}\ }\textbf {\bibinfo {volume} {10}},\
  \bibinfo {pages} {095010} (\bibinfo {year} {2008})},\ \Eprint
  {http://arxiv.org/abs/0802.1842} {arXiv:0802.1842} \BibitemShut {NoStop}%
\bibitem [{\citenamefont {Hertzberg}\ \emph {et~al.}(2010)\citenamefont
  {Hertzberg}, \citenamefont {Rocheleau}, \citenamefont {Ndukum}, \citenamefont
  {Savva}, \citenamefont {Clerk},\ and\ \citenamefont
  {Schwab}}]{Hertzberg2010}%
  \BibitemOpen
  \bibfield  {author} {\bibinfo {author} {\bibfnamefont {J.~B.}\ \bibnamefont
  {Hertzberg}}, \bibinfo {author} {\bibfnamefont {T.}~\bibnamefont
  {Rocheleau}}, \bibinfo {author} {\bibfnamefont {T.}~\bibnamefont {Ndukum}},
  \bibinfo {author} {\bibfnamefont {M.}~\bibnamefont {Savva}}, \bibinfo
  {author} {\bibfnamefont {A.~A.}\ \bibnamefont {Clerk}}, \ and\ \bibinfo
  {author} {\bibfnamefont {K.~C.}\ \bibnamefont {Schwab}},\ }\href {\doibase
  10.1038/nphys1479} {\bibfield  {journal} {\bibinfo  {journal} {Nature
  Physics}\ }\textbf {\bibinfo {volume} {6}},\ \bibinfo {pages} {213} (\bibinfo
  {year} {2010})}\BibitemShut {NoStop}%
\bibitem [{\citenamefont {Suh}\ \emph {et~al.}(2014)\citenamefont {Suh},
  \citenamefont {Weinstein}, \citenamefont {Lei}, \citenamefont {Wollman},
  \citenamefont {Steinke}, \citenamefont {Meystre}, \citenamefont {Clerk},\
  and\ \citenamefont {Schwab}}]{Suh2014}%
  \BibitemOpen
  \bibfield  {author} {\bibinfo {author} {\bibfnamefont {J.}~\bibnamefont
  {Suh}}, \bibinfo {author} {\bibfnamefont {A.~J.}\ \bibnamefont {Weinstein}},
  \bibinfo {author} {\bibfnamefont {C.~U.}\ \bibnamefont {Lei}}, \bibinfo
  {author} {\bibfnamefont {E.~E.}\ \bibnamefont {Wollman}}, \bibinfo {author}
  {\bibfnamefont {S.~K.}\ \bibnamefont {Steinke}}, \bibinfo {author}
  {\bibfnamefont {P.}~\bibnamefont {Meystre}}, \bibinfo {author} {\bibfnamefont
  {A.~A.}\ \bibnamefont {Clerk}}, \ and\ \bibinfo {author} {\bibfnamefont
  {K.~C.}\ \bibnamefont {Schwab}},\ }\href {\doibase 10.1126/science.1253258}
  {\bibfield  {journal} {\bibinfo  {journal} {Science}\ }\textbf {\bibinfo
  {volume} {344}},\ \bibinfo {pages} {1262} (\bibinfo {year} {2014})},\ \Eprint
  {http://arxiv.org/abs/1312.4084} {arXiv:1312.4084} \BibitemShut {NoStop}%
\bibitem [{\citenamefont {Lecocq}\ \emph {et~al.}(2015)\citenamefont {Lecocq},
  \citenamefont {Clark}, \citenamefont {Simmonds}, \citenamefont {Aumentado},\
  and\ \citenamefont {Teufel}}]{Lecocq2015}%
  \BibitemOpen
  \bibfield  {author} {\bibinfo {author} {\bibfnamefont {F.}~\bibnamefont
  {Lecocq}}, \bibinfo {author} {\bibfnamefont {J.~B.}\ \bibnamefont {Clark}},
  \bibinfo {author} {\bibfnamefont {R.~W.}\ \bibnamefont {Simmonds}}, \bibinfo
  {author} {\bibfnamefont {J.}~\bibnamefont {Aumentado}}, \ and\ \bibinfo
  {author} {\bibfnamefont {J.~D.}\ \bibnamefont {Teufel}},\ }\href {\doibase
  10.1103/PhysRevX.5.041037} {\bibfield  {journal} {\bibinfo  {journal}
  {Physical Review X}\ }\textbf {\bibinfo {volume} {5}},\ \bibinfo {pages}
  {041037} (\bibinfo {year} {2015})},\ \Eprint
  {http://arxiv.org/abs/1509.01629v1} {arXiv:1509.01629v1} \BibitemShut
  {NoStop}%
\bibitem [{\citenamefont {Lei}\ \emph {et~al.}(2016)\citenamefont {Lei},
  \citenamefont {Weinstein}, \citenamefont {Suh}, \citenamefont {Wollman},
  \citenamefont {Kronwald}, \citenamefont {Marquardt}, \citenamefont {Clerk},\
  and\ \citenamefont {Schwab}}]{Lei2016}%
  \BibitemOpen
  \bibfield  {author} {\bibinfo {author} {\bibfnamefont {C.~U.}\ \bibnamefont
  {Lei}}, \bibinfo {author} {\bibfnamefont {A.~J.}\ \bibnamefont {Weinstein}},
  \bibinfo {author} {\bibfnamefont {J.}~\bibnamefont {Suh}}, \bibinfo {author}
  {\bibfnamefont {E.~E.}\ \bibnamefont {Wollman}}, \bibinfo {author}
  {\bibfnamefont {A.}~\bibnamefont {Kronwald}}, \bibinfo {author}
  {\bibfnamefont {F.}~\bibnamefont {Marquardt}}, \bibinfo {author}
  {\bibfnamefont {A.~A.}\ \bibnamefont {Clerk}}, \ and\ \bibinfo {author}
  {\bibfnamefont {K.~C.}\ \bibnamefont {Schwab}},\ }\href {\doibase
  10.1103/PhysRevLett.117.100801} {\bibfield  {journal} {\bibinfo  {journal}
  {Physical Review Letters}\ }\textbf {\bibinfo {volume} {117}},\ \bibinfo
  {pages} {100801} (\bibinfo {year} {2016})},\ \Eprint
  {http://arxiv.org/abs/1605.08148} {arXiv:1605.08148} \BibitemShut {NoStop}%
\bibitem [{\citenamefont {Polzik}\ and\ \citenamefont
  {Hammerer}(2014)}]{Polzik2014}%
  \BibitemOpen
  \bibfield  {author} {\bibinfo {author} {\bibfnamefont {E.~S.}\ \bibnamefont
  {Polzik}}\ and\ \bibinfo {author} {\bibfnamefont {K.}~\bibnamefont
  {Hammerer}},\ }\href {\doibase 10.1002/andp.201400099} {\bibfield  {journal}
  {\bibinfo  {journal} {Annalen der Physik}\ }\textbf {\bibinfo {volume}
  {527}},\ \bibinfo {pages} {A15} (\bibinfo {year} {2014})},\ \Eprint
  {http://arxiv.org/abs/1405.3067} {arXiv:1405.3067} \BibitemShut {NoStop}%
\bibitem [{\citenamefont {Malz}\ and\ \citenamefont
  {Nunnenkamp}(2016)}]{Malz2016}%
  \BibitemOpen
  \bibfield  {author} {\bibinfo {author} {\bibfnamefont {D.}~\bibnamefont
  {Malz}}\ and\ \bibinfo {author} {\bibfnamefont {A.}~\bibnamefont
  {Nunnenkamp}},\ }\href {\doibase 10.1103/PhysRevA.94.023803} {\bibfield
  {journal} {\bibinfo  {journal} {Physical Review A}\ }\textbf {\bibinfo
  {volume} {94}},\ \bibinfo {pages} {023803} (\bibinfo {year} {2016})},\
  \Eprint {http://arxiv.org/abs/1605.04749} {arXiv:1605.04749} \BibitemShut
  {NoStop}%
\bibitem [{\citenamefont {Aspelmeyer}\ \emph {et~al.}(2014)\citenamefont
  {Aspelmeyer}, \citenamefont {Kippenberg},\ and\ \citenamefont
  {Marquardt}}]{Aspelmeyer2014}%
  \BibitemOpen
  \bibfield  {author} {\bibinfo {author} {\bibfnamefont {M.}~\bibnamefont
  {Aspelmeyer}}, \bibinfo {author} {\bibfnamefont {T.~J.}\ \bibnamefont
  {Kippenberg}}, \ and\ \bibinfo {author} {\bibfnamefont {F.}~\bibnamefont
  {Marquardt}},\ }\href {\doibase 10.1103/RevModPhys.86.1391} {\bibfield
  {journal} {\bibinfo  {journal} {Reviews of Modern Physics}\ }\textbf
  {\bibinfo {volume} {86}},\ \bibinfo {pages} {1391} (\bibinfo {year}
  {2014})}\BibitemShut {NoStop}%
\bibitem [{\citenamefont {Gardiner}\ and\ \citenamefont
  {Zoller}(2004)}]{gardiner2004quantum}%
  \BibitemOpen
  \bibfield  {author} {\bibinfo {author} {\bibfnamefont {C.}~\bibnamefont
  {Gardiner}}\ and\ \bibinfo {author} {\bibfnamefont {P.}~\bibnamefont
  {Zoller}},\ }\href {https://books.google.de/books?id=a\_xsT8oGhdgC} {\emph
  {\bibinfo {title} {{Quantum Noise: A Handbook of Markovian and Non-Markovian
  Quantum Stochastic Methods with Applications to Quantum Optics}}}},\ Springer
  Series in Synergetics\ (\bibinfo  {publisher} {Springer},\ \bibinfo {year}
  {2004})\BibitemShut {NoStop}%
\bibitem [{\citenamefont {Weinstein}\ \emph {et~al.}(2014)\citenamefont
  {Weinstein}, \citenamefont {Lei}, \citenamefont {Wollman}, \citenamefont
  {Suh}, \citenamefont {Metelmann}, \citenamefont {Clerk},\ and\ \citenamefont
  {Schwab}}]{Weinstein2014}%
  \BibitemOpen
  \bibfield  {author} {\bibinfo {author} {\bibfnamefont {A.~J.}\ \bibnamefont
  {Weinstein}}, \bibinfo {author} {\bibfnamefont {C.~U.}\ \bibnamefont {Lei}},
  \bibinfo {author} {\bibfnamefont {E.~E.}\ \bibnamefont {Wollman}}, \bibinfo
  {author} {\bibfnamefont {J.}~\bibnamefont {Suh}}, \bibinfo {author}
  {\bibfnamefont {A.}~\bibnamefont {Metelmann}}, \bibinfo {author}
  {\bibfnamefont {A.~A.}\ \bibnamefont {Clerk}}, \ and\ \bibinfo {author}
  {\bibfnamefont {K.~C.}\ \bibnamefont {Schwab}},\ }\href {\doibase
  10.1103/PhysRevX.4.041003} {\bibfield  {journal} {\bibinfo  {journal}
  {Physical Review X}\ }\textbf {\bibinfo {volume} {4}},\ \bibinfo {pages}
  {041003} (\bibinfo {year} {2014})},\ \Eprint {http://arxiv.org/abs/1404.3242}
  {arXiv:1404.3242} \BibitemShut {NoStop}%
\bibitem [{\citenamefont {Mari}\ and\ \citenamefont {Eisert}(2009)}]{Mari2009}%
  \BibitemOpen
  \bibfield  {author} {\bibinfo {author} {\bibfnamefont {A.}~\bibnamefont
  {Mari}}\ and\ \bibinfo {author} {\bibfnamefont {J.}~\bibnamefont {Eisert}},\
  }\href {\doibase 10.1103/PhysRevLett.103.213603} {\bibfield  {journal}
  {\bibinfo  {journal} {Physical Review Letters}\ }\textbf {\bibinfo {volume}
  {103}},\ \bibinfo {pages} {213603} (\bibinfo {year} {2009})},\ \Eprint
  {http://arxiv.org/abs/0911.0433} {arXiv:0911.0433} \BibitemShut {NoStop}%
\bibitem [{\citenamefont {Kronwald}\ \emph {et~al.}(2013)\citenamefont
  {Kronwald}, \citenamefont {Marquardt},\ and\ \citenamefont
  {Clerk}}]{Kronwald2013}%
  \BibitemOpen
  \bibfield  {author} {\bibinfo {author} {\bibfnamefont {A.}~\bibnamefont
  {Kronwald}}, \bibinfo {author} {\bibfnamefont {F.}~\bibnamefont {Marquardt}},
  \ and\ \bibinfo {author} {\bibfnamefont {A.~A.}\ \bibnamefont {Clerk}},\
  }\href {\doibase 10.1103/PhysRevA.88.063833} {\bibfield  {journal} {\bibinfo
  {journal} {Physical Review A}\ }\textbf {\bibinfo {volume} {88}},\ \bibinfo
  {pages} {063833} (\bibinfo {year} {2013})}\BibitemShut {NoStop}%
\bibitem [{\citenamefont {Pirkkalainen}\ \emph {et~al.}(2015)\citenamefont
  {Pirkkalainen}, \citenamefont {Damsk\"{a}gg}, \citenamefont {Brandt},
  \citenamefont {Massel},\ and\ \citenamefont
  {Sillanp\"{a}\"{a}}}]{Pirkkalainen2015}%
  \BibitemOpen
  \bibfield  {author} {\bibinfo {author} {\bibfnamefont {J.-M.}\ \bibnamefont
  {Pirkkalainen}}, \bibinfo {author} {\bibfnamefont {E.}~\bibnamefont
  {Damsk\"{a}gg}}, \bibinfo {author} {\bibfnamefont {M.}~\bibnamefont
  {Brandt}}, \bibinfo {author} {\bibfnamefont {F.}~\bibnamefont {Massel}}, \
  and\ \bibinfo {author} {\bibfnamefont {M.~A.}\ \bibnamefont
  {Sillanp\"{a}\"{a}}},\ }\href {\doibase 10.1103/PhysRevLett.115.243601}
  {\bibfield  {journal} {\bibinfo  {journal} {Physical Review Letters}\
  }\textbf {\bibinfo {volume} {115}},\ \bibinfo {pages} {243601} (\bibinfo
  {year} {2015})},\ \Eprint {http://arxiv.org/abs/1507.04209}
  {arXiv:1507.04209} \BibitemShut {NoStop}%
\bibitem [{\citenamefont {Wollman}\ \emph {et~al.}(2015)\citenamefont
  {Wollman}, \citenamefont {Lei}, \citenamefont {Weinstein}, \citenamefont
  {Suh}, \citenamefont {Kronwald}, \citenamefont {Marquardt}, \citenamefont
  {Clerk},\ and\ \citenamefont {Schwab}}]{Wollman2015}%
  \BibitemOpen
  \bibfield  {author} {\bibinfo {author} {\bibfnamefont {E.~E.}\ \bibnamefont
  {Wollman}}, \bibinfo {author} {\bibfnamefont {C.~U.}\ \bibnamefont {Lei}},
  \bibinfo {author} {\bibfnamefont {A.~J.}\ \bibnamefont {Weinstein}}, \bibinfo
  {author} {\bibfnamefont {J.}~\bibnamefont {Suh}}, \bibinfo {author}
  {\bibfnamefont {A.}~\bibnamefont {Kronwald}}, \bibinfo {author}
  {\bibfnamefont {F.}~\bibnamefont {Marquardt}}, \bibinfo {author}
  {\bibfnamefont {A.~A.}\ \bibnamefont {Clerk}}, \ and\ \bibinfo {author}
  {\bibfnamefont {K.~C.}\ \bibnamefont {Schwab}},\ }\href {\doibase
  10.1126/science.aac5138} {\bibfield  {journal} {\bibinfo  {journal}
  {Science}\ }\textbf {\bibinfo {volume} {349}},\ \bibinfo {pages} {952}
  (\bibinfo {year} {2015})},\ \Eprint {http://arxiv.org/abs/1507.01662}
  {arXiv:1507.01662} \BibitemShut {NoStop}%
\bibitem [{\citenamefont {Woolley}\ and\ \citenamefont
  {Clerk}(2013)}]{Woolley2013}%
  \BibitemOpen
  \bibfield  {author} {\bibinfo {author} {\bibfnamefont {M.~J.}\ \bibnamefont
  {Woolley}}\ and\ \bibinfo {author} {\bibfnamefont {A.~A.}\ \bibnamefont
  {Clerk}},\ }\href {\doibase 10.1103/PhysRevA.87.063846} {\bibfield  {journal}
  {\bibinfo  {journal} {Physical Review A}\ }\textbf {\bibinfo {volume} {87}},\
  \bibinfo {pages} {063846} (\bibinfo {year} {2013})},\ \Eprint
  {http://arxiv.org/abs/1304.4059} {arXiv:1304.4059} \BibitemShut {NoStop}%
\bibitem [{\citenamefont {Ockeloen-Korppi}\ \emph {et~al.}(2016)\citenamefont
  {Ockeloen-Korppi}, \citenamefont {Damsk\"{a}gg}, \citenamefont
  {Pirkkalainen}, \citenamefont {Clerk}, \citenamefont {Woolley},\ and\
  \citenamefont {Sillanp\"{a}\"{a}}}]{Ockeloen-Korppi2016}%
  \BibitemOpen
  \bibfield  {author} {\bibinfo {author} {\bibfnamefont {C.~F.}\ \bibnamefont
  {Ockeloen-Korppi}}, \bibinfo {author} {\bibfnamefont {E.}~\bibnamefont
  {Damsk\"{a}gg}}, \bibinfo {author} {\bibfnamefont {J.-M.}\ \bibnamefont
  {Pirkkalainen}}, \bibinfo {author} {\bibfnamefont {A.~A.}\ \bibnamefont
  {Clerk}}, \bibinfo {author} {\bibfnamefont {M.~J.}\ \bibnamefont {Woolley}},
  \ and\ \bibinfo {author} {\bibfnamefont {M.~A.}\ \bibnamefont
  {Sillanp\"{a}\"{a}}},\ }\href {\doibase 10.1103/PhysRevLett.117.140401}
  {\bibfield  {journal} {\bibinfo  {journal} {Physical Review Letters}\
  }\textbf {\bibinfo {volume} {117}},\ \bibinfo {pages} {140401} (\bibinfo
  {year} {2016})}\BibitemShut {NoStop}%
\bibitem [{\citenamefont {Walls}\ and\ \citenamefont
  {Milburn}(2008)}]{Walls2008}%
  \BibitemOpen
  \bibfield  {author} {\bibinfo {author} {\bibfnamefont {D.}~\bibnamefont
  {Walls}}\ and\ \bibinfo {author} {\bibfnamefont {G.~J.}\ \bibnamefont
  {Milburn}},\ }\href {\doibase 10.1007/978-3-540-28574-8} {\emph {\bibinfo
  {title} {{Quantum Optics}}}},\ edited by\ \bibinfo {editor} {\bibfnamefont
  {D.}~\bibnamefont {Walls}}\ and\ \bibinfo {editor} {\bibfnamefont {G.~J.}\
  \bibnamefont {Milburn}}\ (\bibinfo  {publisher} {Springer Berlin
  Heidelberg},\ \bibinfo {address} {Berlin, Heidelberg},\ \bibinfo {year}
  {2008})\BibitemShut {NoStop}%
\bibitem [{\citenamefont {Tsang}\ and\ \citenamefont
  {Caves}(2012)}]{Tsang2012}%
  \BibitemOpen
  \bibfield  {author} {\bibinfo {author} {\bibfnamefont {M.}~\bibnamefont
  {Tsang}}\ and\ \bibinfo {author} {\bibfnamefont {C.~M.}\ \bibnamefont
  {Caves}},\ }\href {\doibase 10.1103/PhysRevX.2.031016} {\bibfield  {journal}
  {\bibinfo  {journal} {Physical Review X}\ }\textbf {\bibinfo {volume} {2}},\
  \bibinfo {pages} {031016} (\bibinfo {year} {2012})}\BibitemShut {NoStop}%
\end{thebibliography}%
\end{document}